\documentclass[a4paper, pra, twocolumn,superscriptaddress,showpacs]{revtex4-1}

\usepackage{amssymb}
\usepackage{amsmath}
\usepackage{epsfig}
\usepackage{color}
\usepackage{graphics, graphicx}
\usepackage{bbold}
\usepackage{psfrag}
\usepackage{mathcomp}
\usepackage{subfigure}
\usepackage{verbatim}
\usepackage{color}
\usepackage[colorlinks,citecolor=blue]{hyperref}

\begin{document}

\date{\today}
\title{The dressed molecules theory for the quasi-two-dimensional quantum anomaly}
\author{Fan Wu }

\thanks{These authors contributed equally to this work.}

\affiliation{Fujian Key Laboratory of Quantum Information and Quantum\\
	Optics, College of Physics and Information Engineering, Fuzhou University,
	Fuzhou, Fujian, 350108, China}

\author{Jian-shen Hu }

\thanks{These authors contributed equally to this work.}
\affiliation{Department of Physics and State Key Laboratory of Low-Dimensional Quantum Physics, Tsinghua University, Beijing 100084, China}

\author{Lianyi He}
\affiliation{Department of Physics and State Key Laboratory of Low-Dimensional Quantum Physics, Tsinghua University, Beijing 100084, China}

\begin{abstract}
In this work, the dressed molecules theory is used to describe the two-dimensional quantum anomaly of breathing mode in the recent experimental system\cite{Holten2018,Peppler2018}. With the aid of a beyond mean-field, Gaussian pair fluctuation theory, we employ the dressed molecules states to characterize the axial excited states and the Feshbach molecular states, and propose a complete low energy effective theory. We show that, in the whole crossover from a Bardeen-Cooper-Schrieffer (BCS) superfluid to a Bose-Einstein condensate (BEC), our theory can describe the two-dimensional experimental systems precisely in low energe region. We explain the puzzling experimental observations of the smaller than expected breathing mode frequency perfectly. Our establishment of the dressed molecules theory for 2D fermions is crucial to understand the conformal anomaly in the quasi low-dimensional quantum systems.
\end{abstract}
\maketitle

\section{Introduction}
The study of the conformal anomaly in the low-dimensional systems has received considerable attention in recent years. Due to the quantum effect, the symmetry of the classical theory can be destroyed, this is called quantum anomaly. One of the best-known anomaly is the conformal anomaly, associated with the violation of scale invariance by quantum corrections or quantified in renormalization. Since the renormalization or the quantum corrections introduces a distance scale, the classically scale-invariant symmetry has been broken. This is ubiquitous in quantum field theory, such as Quantum Electrodynamics (QED), Quantum Chromodynamics (QCD), and the Gross-Neveu model \cite{QA}. 

In the two-dimensional system, these quantum anomaly can be observed by measuring the breathing mode of two-component interacting Fermi gases \cite{Hofmann}. In the past, many articles have predicted the observed quantum abnormal breathing mode from the theoretical calculation and the quantum Monte Carlo simulations\cite{Randeria1989,Berezinskii1972,Kosterlitz1973,Levinsen2015,Turlapov2017,Makhalov2014,Martiyanov2016,Fenech2016,Boettcher2016,Frohlich2011,Sommer2012,Zhang2012,Ries2015,Murthy2015,Vogt2012,Holten2018,Peppler2018}. However, the latest two experimental results show that even in the deep two-dimensional region, the predicted results are still quite different from the experimental data\cite{Holten2018,Peppler2018}. The observed frequency is far less than the well-established theoretical prediction in the strongly interaction region\cite{JH,YuZH}.

The purpose of this work is to provide a complete theory of the quasi-two-dimensional quantum anomaly for explaining the experimental observations of the quantum anomaly of the breathing mode frequency. In the previous work, we established a minimal model to describe ultracold interacting fermions confined in two dimensions and solved it accurately at zero temperature with the help of existing Auxiliary-field quantum Monte Carlo (AFQMC) results\cite{WF}. In the Bardeen-Cooper-Schrieffer (BCS) superfluid region, this minimal model resolve in a satisfactory way the puzzling experimental observations of the smaller than expected equations of state and breathing mode frequency. But the agreement between the theoretical calculation and experiment datas becomes worse in Bose-Einstein condensate (BEC) region, suggesting the inadequacy of our theory towards the limit of a BEC. This makes it necessary to put forward a more complete theory for describing the two-dimensional interaction ultracold fermions in the experimental system with the whole BCS-BEC crossover. While, a more complete theory for explaining experimental observations needs to meet these following conditions. First, this theory must be able to characterize the quantum anomaly of the breathing mode frequency, so it must be a beyond mean-field theory. Secondly, because of the strong interaction region, the energy of the z-axis potential can not always be much larger than other energy indices scales.This makes that the axial excited states' fermions effect cannot be neglected. Finally, in the BEC region, the properties of the Feshbach molecular states will be dominant. 

In this work, by using the Gaussian pair fluctuation theory to describe the quantum properties of the system, and employing the dressed molecules states\cite{WZ} to characterize the axial excited states and the Feshbach molecular states, we develop a more complete theory . We show that, in the whole crossover from a Bardeen-Cooper-Schrieffer (BCS) superfluid to a Bose-Einstein condensate (BEC), our theory can describe the two-dimensional experimental systems precisely in low energe region. The paper is organized as follows: In Sec. \ref{model}, we establish the dressed molecules model by fitting the two-body parameters with the two-channel model of three dimensional experimental system. We also verify the completeness of this model in the whole BCS-BEC crossover.  In Sec. \ref{GPF},  we formulate the many-body theory of the dressed molecules model, and give the calculation method of breathing mode frequency.  In Sec. \ref{res}, we show the many-body calculation results. We summarize in Sec. \ref{Sum}.

\section{Dressed molecules model}\label{model}

\subsection{System description}
In order to obtain a complete theory describing the quasi two dimensional quantum anomaly, we need to get an effective model which can completely characterise the few body physics of the quasi two dimensional system. So fist, we focus on the two body problem in the three dimensional two-component Fermi gases with s-wave interaction by an one-dimensional z directional harmonic trap. This two-body Hamiltonionian is taken by the two-channel form:
\begin{align}
H=&H_0+H_{\rm bf}+H_{\rm int}\nonumber \\
H_0 =& \sum_{\sigma = \uparrow, \downarrow} \int d^3 \boldsymbol{r} \psi_{\sigma}^\dagger
\left( - \frac{\hbar^2 \nabla^2}{2m_f} + \frac{1}{2} m_f \omega_z^2 z^2 \right) \psi_{\sigma}
\nonumber \\
&+ \int d^3\boldsymbol{r} \phi^\dagger
\left( - \frac{\hbar^2 \nabla^2}{4m_f} + m_f \omega_z^2 z^2 + {\bar \nu}_b \right) \phi,
\nonumber \\
H_{\rm bf} =& {\bar g}_b \int d^3\boldsymbol{r}
\left( \phi^\dagger \psi_\downarrow \psi_\uparrow + {\rm H.C.} \right)
\nonumber \\
H_{\rm int} =&
{\bar U}_b \int d^3\boldsymbol{r} \psi_\uparrow^\dagger \psi_\downarrow^\dagger
\psi_\downarrow \psi_\uparrow ,
\end{align}
here $\psi_{\sigma}$ and $\phi$ are the atomic and molecular field operator, $\omega_z$ is the trap frequency of the z directional harmonic trap, ${\bar \nu}_b$, ${\bar g}_b$ and ${\bar U}_b$ are corresponding to the bare detuning, the bare atom-molecule coupling constant and the bare background scattering amplitude. These bare scattering parameters are related to the physical ones (just like ${\bar U}_p$) via the renormalization relations \cite{QC}:
\begin{align}
&U_{c}^{-1} = - \int \frac{d^3\boldsymbol{k}}{(2\pi^3)} \frac{1}{2 {\bar\epsilon}_{\boldsymbol{k}}},
\quad
\Gamma^{-1} = 1 + \frac{{ U}_p}{U_c},
\nonumber \\
&{ U}_p = \Gamma^{-1} {\bar U}_b, \quad { g}_p = \Gamma^{-1} {\bar g}_b, \quad
{ \nu}_p = {\bar \nu}_b + \Gamma \frac{{ g}_p^2}{U_c}
\label{eqn:3DR}.
\end{align}
The physical parameters are determined from the scattering data as
\begin{align}
{ U}_p =&\frac{4 \pi \hbar^2 a_{\rm bg}}{m_f},
\hspace{0.2cm}{ g}_p = \sqrt{\frac{4 \pi \hbar^2 \mu_{\rm co} W |a_{\rm bg}|}{m_f}},\nonumber\\
{ \nu}_p =& \mu_{\rm co} (B-B_0),
\end{align}
where $a_{\rm bg}$is the background scattering length, $W$ is the resonance width, $\mu_{\rm co}$ is the difference in the magnetic moments of the closed and open channels, and $B_0$ is the resonance position.

To better understanding this Hamiltonionian, we use harmonic modes and plane waves to expand the field operators in the trapped z-direction and the untrapped x-y plane. Here we just consider the center of mass is zero case as it is decoupled from the relative momentum in the two-body case. The Hamiltonionian is taken by:
\begin{align}
H_0 =& \sum_{m, {\bf k}, \sigma}
\epsilon_{m,{\bf k}}
c_{m, {\bf k},\sigma}^\dagger c_{m, {\bf k},\sigma}+\nu_b b_{0}^\dagger b_{0}
\nonumber \\
H_{\rm bf}  =&  g_b \sum_{m,n,  {\bf k}} \gamma_{mn}
\bigg( b_{0}^\dagger c_{m, -{\bf k}, \downarrow}
c_{n, {\bf k}, \uparrow}
+ {\rm H.C.} \bigg)\nonumber\\
H_{\rm int} =& U_b \sum_{m,n, {\bf k},m^\prime,n^\prime, {\bf k}^\prime}
c_{m, {\bf k},\uparrow}^\dagger
c_{n, -{\bf k},\downarrow}^\dagger c_{n^\prime, -{\bf k}^\prime,\downarrow}
c_{m^\prime, {\bf k}^\prime,\uparrow}
\label{eqn:3DM}.
\end{align}

Here the Hamiltonion is the dimensionless Hamiltonionian with the energy unit $E_0=\hbar\omega_z$ and the length unit $a_t  = \sqrt{\hbar/(m_f \omega_z)}$. And The bosonic field $b_0$ represents the molecular state with transverse momentum $k = 0$ and axial harmonic mode $m = 0$. The atom relative energy $\epsilon_{m,{\bf k}}=1/4+m+k_x^2+k_y^2$. And for the coefficient $\gamma_{mn}$, when $m+n$ is odd $\gamma_{mn}=0$, when $m+n$ is even,
\begin{align}
\gamma_{mn}=\frac{(-1)^{(m-n)/2}}{(2\pi^3)^{1/4}\sqrt{m!n!}}\Gamma(\frac{m+n+1}{2}).
\end{align}
And to the bare parameter, $g_b = {\bar g}_b a_t^{-3/2} / ({\hbar \omega_z})$, $U_b = {\bar U}_b a_t^{-3} / ({\hbar \omega_z})$, $\nu_b = {\bar \nu}_b / ({\hbar \omega_z})$.

We can find that, in the Hamiltonionian (\ref{eqn:3DM}), there are three types of states, fermions in the axial ground state, fermions in axial excited states, and Feshbach molecular states. In order to fully characterize these states, we use the two-dimensional dressed molecules model. It is a two channel model in which the particles in the open channel are fermions, representing the fermions in the axial ground state, and the particles in the closed channel are the dressed bosons, representing the fermions in axial excited states, and the Feshbach molecules\cite{WZ,LMD1,LMD2,WZ1,WZ2,WY1}. This effective Hamiltonionian is also written in dimensionless form with length unit $a_t$ and energy unit $\hbar\omega_z$:
\begin{align}
\label{eff-H}
H_{\rm eff}=&
\sum_{{\bf k},\sigma} \epsilon_{\bf k} a_{{\bf k},\sigma}^\dagger a_{{\bf k}, \sigma}
+ \sum_{\bf q} \left(\delta_b + \epsilon_{\bf q}/2 \right) d_{\bf q}^\dagger d_{\bf q}\nonumber\\
&+ \alpha_b \sum_{{\bf k},{\bf q}} \left( d_{\bf q}^\dagger a_{{\bf k}+{\bf q}/2,\uparrow}
a_{-{\bf k}+{\bf q}/2,\downarrow}  + {\rm H.C.} \right)\nonumber\\
&+ V_b \sum_{{\bf k},{\bf k}^\prime,{\bf q}}
a_{{\bf k}+{\bf q}/2,\uparrow}^\dagger a_{-{\bf k}+{\bf q}/2,\downarrow}^\dagger
a_{-{\bf k}^\prime+{\bf q}/2, \downarrow} a_{{\bf k}^\prime+{\bf q}/2, \uparrow}.
\end{align}
Here, $\epsilon_{\bf k}=k^2/2$, and the $\delta_b, \alpha_b, V_b$ are the three bare scattering parameters. And $a_{{\bf k},\sigma}^\dagger(a_{{\bf k}, \sigma})$ are fermionic creation (annihilation) operator and $d_{\bf q}^\dagger(d_{\bf q})$ are bosonic creation (annihilation) operator. Here, the dressed molecules are structureless, because all short-range details associated with the fermions in axial excited states and the Feshbach molecules are irrelevant in the low-energy region.

The bare scattering parameters can be linked to physical ones by use the 2D renormalization analogous to Eq.(\ref{eqn:3DR}):
\begin{align}
&V_{c}^{-1} = - \int \frac{d^2\boldsymbol{k}}{(4\pi^2)} \frac{1}{2 {\bar\epsilon}_{\boldsymbol{k}}+1},
\quad
\Omega^{-1} = 1 + \frac{{ V}_p}{V_c},
\nonumber \\
&{ V}_p = \Omega^{-1} { V}_b, \quad { \alpha}_p = \Omega^{-1} {\alpha}_b, \quad
{ \delta}_p = { \delta}_b + \Omega \frac{{\alpha}_p^2}{V_c}
\label{eqn:2DR}.
\end{align}

\subsection{Parameter fitting}
We're going to fix the physical scattering parameters in the dressed molecules model. To ensure the low energy efficiency of the dressed molecules model, we use the $T$ matrix as the benchmark to do the parameter fitting.

From the Hamiltonion (\ref{eqn:3DM}), we can get the $T$ matrix of the three dimensional system:
\begin{align}
T_{\rm 3D}(E_{\rm 3D})^{-1}=\sqrt{2\pi}\left\{\left[ U_p - \frac{g_p^2}{\nu_p - E_{\rm3D} }\right]^{-1}-S_p(E_{\rm3D})\right\}\label{eq:T3D}.
\end{align}

Here, $S_p(E_{\rm3D})$ is the two particle bubble function of the system. We can get it by solving the two-body band state equation with the general two-body state ansatz involving the atoms and the molecule.
And the two particle bubble function is
\begin{align}
S_p(E_{\rm3D}) & \equiv  \sum_{m,n,{\bf k}} \gamma_{mn}^2
\left[
{\cal E}_{m,n, {\bf k}}^{-1}
+
\frac{1}{2\epsilon_{\bf k}}
\right]\nonumber\\
&=\frac{-1}{4\sqrt{2}\pi}\int_0^{+\infty}ds[\frac{\Gamma(s+1/4-E_{\rm3D}/2)}{\Gamma(s+3/4-E_{\rm3D}/2)}-1/\sqrt{s}].\label{eqn:etaseqn}
\end{align}
Here ${\cal E}_{m,n, {\bf k}} = E_{3D} - k^2 - 1-m-n$, $E_{\rm3D}$ is the energy of the two body system in the Hamiltonion (\ref{eqn:3DM}). This is equivalent to the quasi-2D bubble function calculated by Petrov and Shlyapnikov\cite{Petrov1}, the details are given in the appendix.

Similarly, for the dressed molecular model's Hamiltonion (\ref{eff-H}),
the $T$ matrix and the two particle bubble function are:

\begin{figure}[tbp]
    \includegraphics[width=0.8\linewidth]{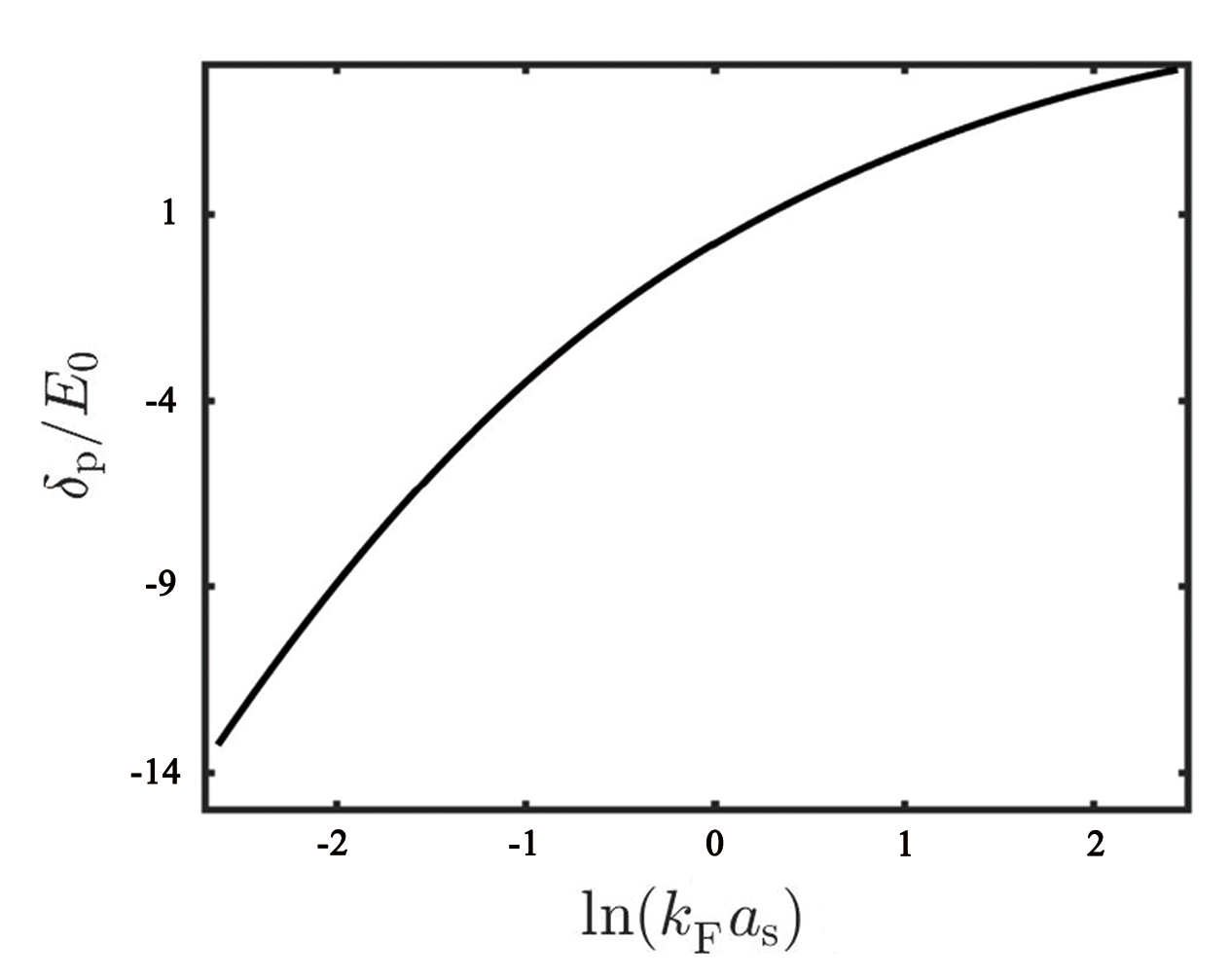}
	\includegraphics[width=0.8\linewidth]{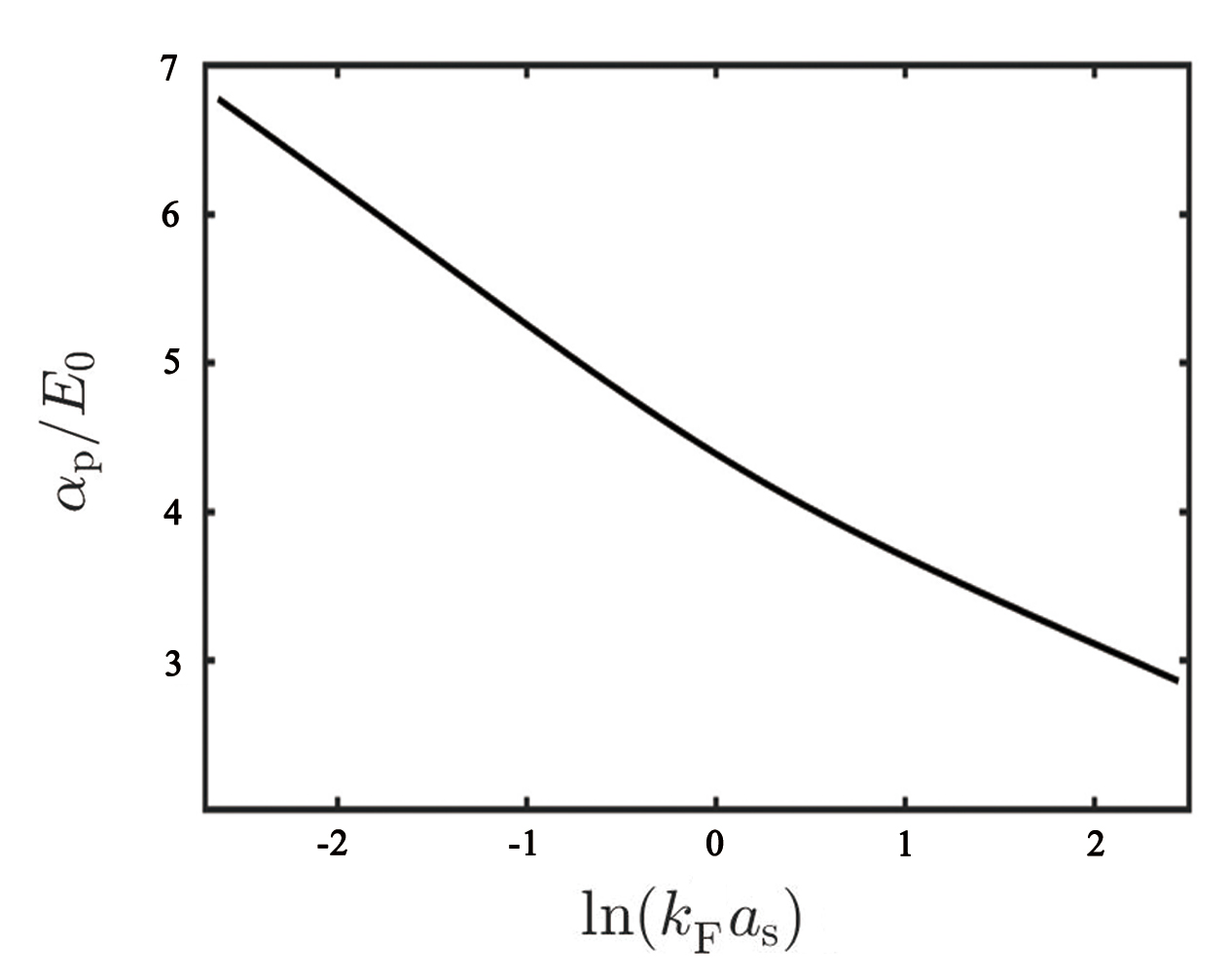}
	\includegraphics[width=0.8\linewidth]{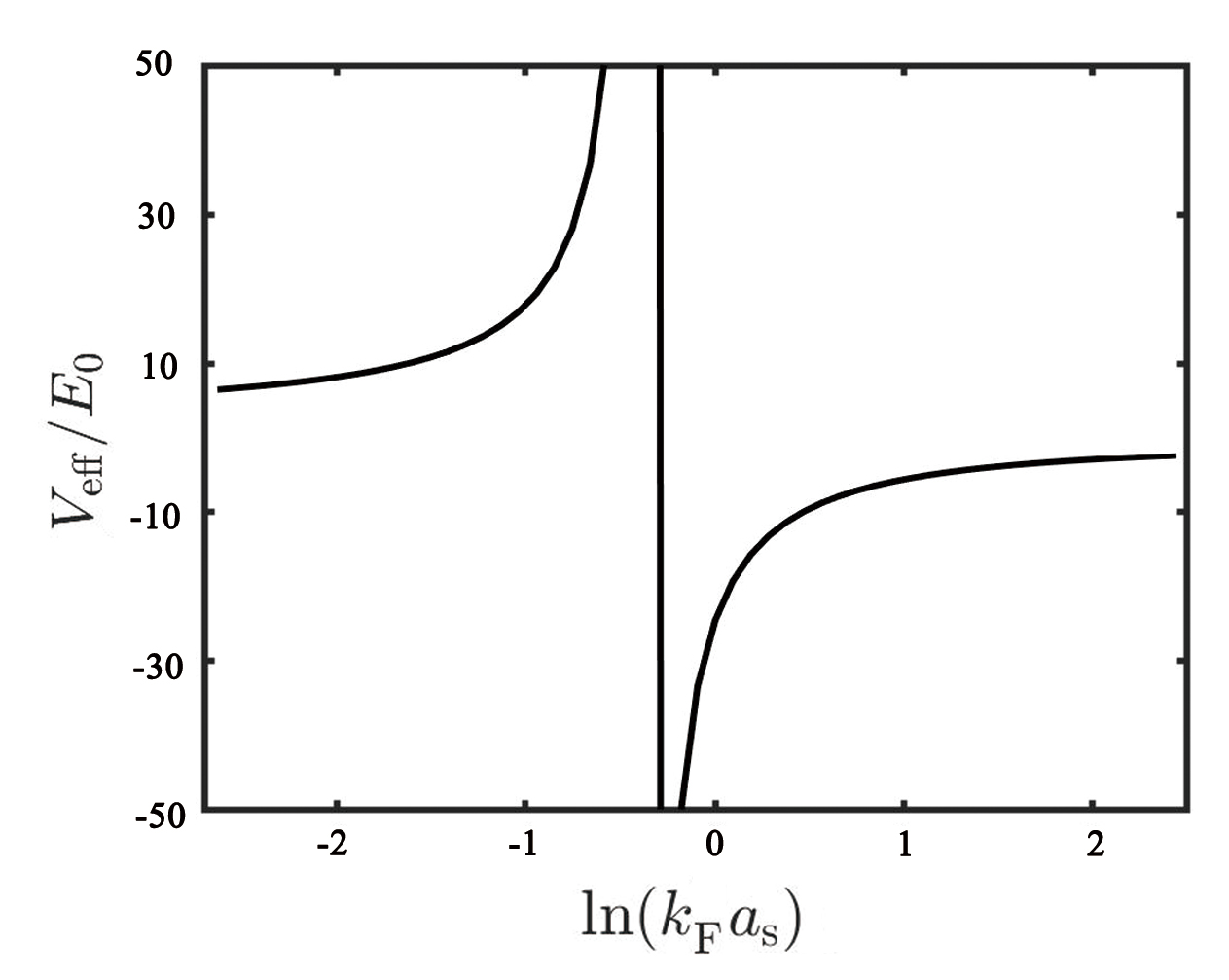}
	\caption{(Color online) The evolution curve of the two body interaction parameters in the dressed molecules model with the change of $\ln(k_{F}a_{s})$. Here, $\delta_{p}$ is the effective energy difference between open and closed channels, $\alpha_{p}$ is the effective coupling strength between open and closed channels, $V_{eff}$ is the effective interaction strength,  and the unit of the two body interaction parameters $E_{0}=\hbar\omega_z$.}
\label{fig:para}
\end{figure}

\begin{align}
T_{\rm2D}(E_{\rm2D})^{-1}=&\left[ V_p - \frac{\alpha_p^2}{\delta_p - E_{\rm2D} }\right]^{-1}-\sqrt{2\pi}\sigma_p(E_{\rm2D}),\label{eq:T2D}\\
\sigma_{p}(E_{\rm2D}) =& \int \frac{d^{2}{\bf k}}{(2\pi)^{5/2}}
\left[ \frac{1}{E_{\rm2D} - 2 \epsilon_{\bf k}}
+ \frac{1}{1+ 2 \epsilon_{\bf k}} \right]\nonumber\\
=&\frac{\ln(-E_{\rm2D})}{4 \pi \sqrt{2 \pi}}.
\end{align}
Here, $E_{\rm2D}$ is the energy of the two body system in the dressed molecular model.

Now we go to do the parameter fitting.
At first, from the Eq.(\ref{eqn:3DM}) and Eq.(\ref{eff-H}) for the vacuum state, the Energy of the vacuum state in the quasi two dimensional Hamiltonionian is $E_{\rm3D}=1/2$, and for the dressed molecular model, it is $E_{\rm2D}=0$. So we can get $E_{\rm3D}=E_{\rm2D}+1/2$. For convenience, we define the two body energy as $E=E_{\rm2D}$.

Second, we consider $\nu_p \to \infty$ case. We defind $E_b^{\rm inf}$ is the bounding energy in the dressed molecular model with $\nu_p \to \infty$. We can get
\begin{align}
U_p^{-1}=&S_p (E_b^{\rm inf}+1/2)
\end{align}
From this equation we can solve the bounding energy $E_b^{inf}$, and we also have $ V_p^{-1}=\sqrt{2 \pi } \sigma_p(E_b^{\rm inf})$.
We defind $C_p=S_p(E_b^{\rm inf}+1/2) - \sigma_p(E_b^{\rm inf})$, then we can fit the parameter $V_{p}$ with the equation
\begin{align}
V_{p}^{-1}=\sqrt{2 \pi} \left(U_p^{-1} - C_p \right).
\end{align}
Third, we know that, the $T$ matrix has simple poles when $E$ is equal to a two-body bound state of the Hamiltonionian.
By matching the pole of the $T$ matrix in Eq.(\ref{eq:T3D}) and Eq.(\ref{eq:T2D}), we can get:
\begin{align}
\left[ V_p - \frac{\alpha_p^2}{ \delta_p - E_b} \right]^{-1} &= \sqrt{2 \pi } \sigma_p(E_b).
\end{align}
Here, $E_b$ is the solution of equation $U_{{\rm eff}}(E)^{-1}=S_p(E+1/2)$, and $U_{{\rm eff}}(E)=U_p - g_p^2/[\nu_p - (E+1/2) ]$.

By matching the first derivative of $T^{-1}$ around the pole, we can get:
\begin{align}
\frac{\partial}{\partial E_b}V_{{\rm eff}}(E_b)^{-1}
=& \frac{\sqrt{2 \pi }\partial}{\partial E_b}[U_{{\rm eff}}(E_b)^{-1}\nonumber\\
&-S_p(E_b+1/2)+\sigma_p(E_b)]
\end{align}
Here, $V_{{\rm eff}}(E_b)= V_p - \alpha_p^2/(\delta_p - E_b)$. Then we can fit the parameters $\delta_p$ and $\alpha_p$ from the equations:
\begin{align}
\delta_p =& E_b - \frac{\sigma_p(E_b)\Lambda(E_B)}{\frac{\partial}{\partial E_b}\left\{U_{{\rm eff}}(E_b)^{-1}-S_p(E_b+1/2)+\sigma_p(E_b)\right\} },
\nonumber \\
\alpha_p^2 =& \frac{\Lambda(E_B)^2}{ \sqrt{2 \pi} \frac{\partial}{\partial E_b}\left\{U_{{\rm eff}}(E_b)^{-1}-S_p(E_b+1/2)+\sigma_p(E_b)\right\}}.
\end{align}

Here, $\Lambda(E_B)=\left[ 1- \sigma_p(E_b)/( U_p^{-1} - C_p) \right]$.

\begin{figure}[tbp]
\includegraphics[width=1\linewidth]{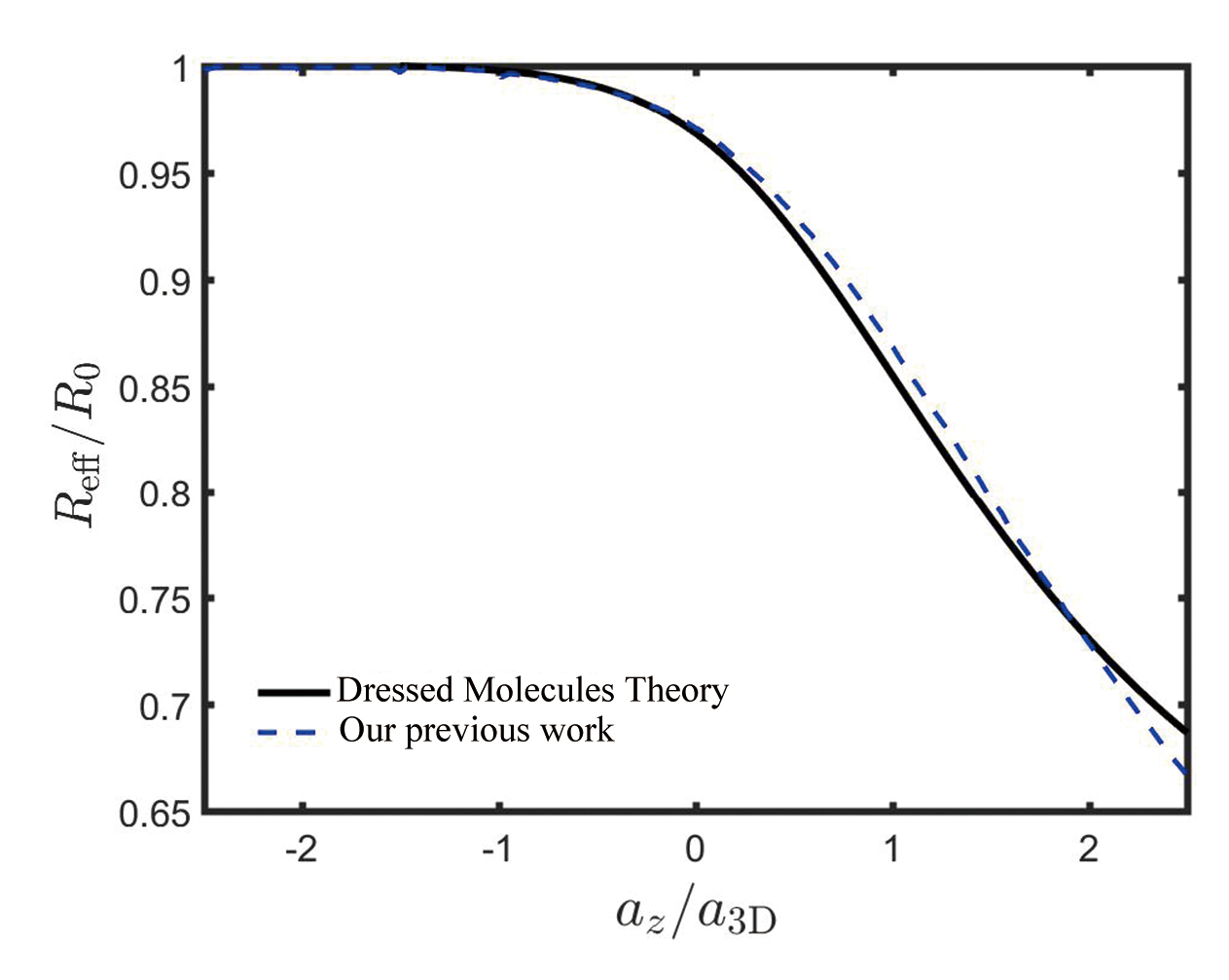}
\caption{(Color online) The effective 2D effective range $R_{\rm eff}$ change with the 3D interaction strength $a_{z}/a_{\rm3D}$. Here, the unit of the $R_{\rm eff}$ is the 2D effective range in the BCS limit $R_{s}^{(0)}=(-\ln2)a_{z}^{2}$, and the black solid line is the result of our dressed molecules theory, and the
Blue dotted line is the date of our previous work\cite{WF}.}\label{fig:Reff}
\end{figure}

In Fig.\ref{fig:para}, we plot the parameters of the dressed molecular model responding to the quasi two dimonsional system\cite{Holten}, here, $V_{\rm eff}$ is the effective interaction which we defined it before, and it will play an important role for the many-body calculation discussed below, the horizontal axis parameters $a_s$ is the effective 2D scattering length $a_s\equiv a_z\sqrt{\pi/\mathcal{B}}\exp{-\sqrt{\pi/2}a_z/a_{\rm3D}}$, and $a_{\rm3D}=(U_p-g_p^2/\nu_p)/(4\pi)$, $\mathcal{B}\sim0.905$\cite{Petrov1}. We also show the effective range of the dressed molecular model in Fig.\ref{fig:Reff}, compare to the effective theory in our previous article \cite{WF1}, the properties of the two curves with effective range are consistent. However, as the parameter approaches to the BEC-limit, the difference between these two curves begins to appear. One hand, it's because of that the model in the effective theory about the previous article is just a the concise model, but the dressed molecular model is the complete model in the low-energy region. On the other hand, in this paper, the three dimensional Hamiltonionian is described in the two-channel form, so in the BEC-side, this description is more physical than that in the previous article\cite{WF1}.

Up to now, we have introduced an effective dressed states 2D Hamiltonionian by matching the two-body physics with the three dimensional systems. This effective theory is the complete model in the low-energy region. By grouping the axial excited states' fermions and Feshbach molecules to define a dressed molecular state, we can keep the correction of the low energy many-body physics. Because of the matching conditions of the pole of the two-body $T$ matrix and the first derivative of $T^{-1}$ around the pole, when the fermionic chemical potential is not far away from the bound state energy, just $|2\mu+|E_b||\ll1$, the $T$ matrix of the three dimensional Hamiltonionian can be well approximated by that of the dressed molecular model. Considering the fact that $|2\mu+|E_b||\leq E_F$ through the BCS-BEC crossover, here $E_F$ is the Fermi energy, and the Fermi energy is proportional to the number density $E_F\propto n$, so for the diluteness and the quasi-2D systems we consider for this paper, this effective dressed states theory is approximately valid.

\section{GAUSSIAN PAIR FLUCTUATION THEORY}\label{GPF}
 We carry out the many-body simulation based on this dressed molecular model. Through Eq.(\ref{eff-H}), in the imaginary time path integral formalism, the partition function ${\cal Z}$ is given by
\begin{align}
Z=&\int \mathfrak{D}[\Psi_{\sigma}^{\dag},\Psi_{\sigma},\Phi^{\dag},\Phi]e^{-{\cal S}_{{\rm eff}}[\Psi_{\sigma}^{\dag},\Psi_{\sigma},\Phi^{\dag},\Phi]},
\end{align}
where $\Psi_{\sigma}$ and $\Phi$ are the real space atomic and molecular field operator corresponding to $a_{{\bf k},\sigma}$ and $d_{\bf q}$ in Eq.(\ref{eff-H}), and $S_{\rm eff}$ is the effective action
\begin{align}
&{\cal S}_{{\rm eff}}[\Psi_{\sigma}^{\dag},\Psi_{\sigma},\Phi^{\dag},\Phi]\nonumber\\
=&\int_{0}^{\beta} d\tau\big\{\int\frac{d^2\boldsymbol{r}}{S} [\Psi^{\dag}_{\sigma}(\boldsymbol{r})\left(\frac{\partial}{\partial\tau}-\mu\right)\Psi_{\sigma}(\boldsymbol{r})\nonumber\\
&
+\Phi^{\dag}(\boldsymbol{r})\left(\frac{\partial}{\partial\tau}-2\mu\right)\Phi(\boldsymbol{r})]+H_{\rm eff}\big\}.
\end{align}

\begin{figure}[tbp]
\includegraphics[width=1\linewidth]{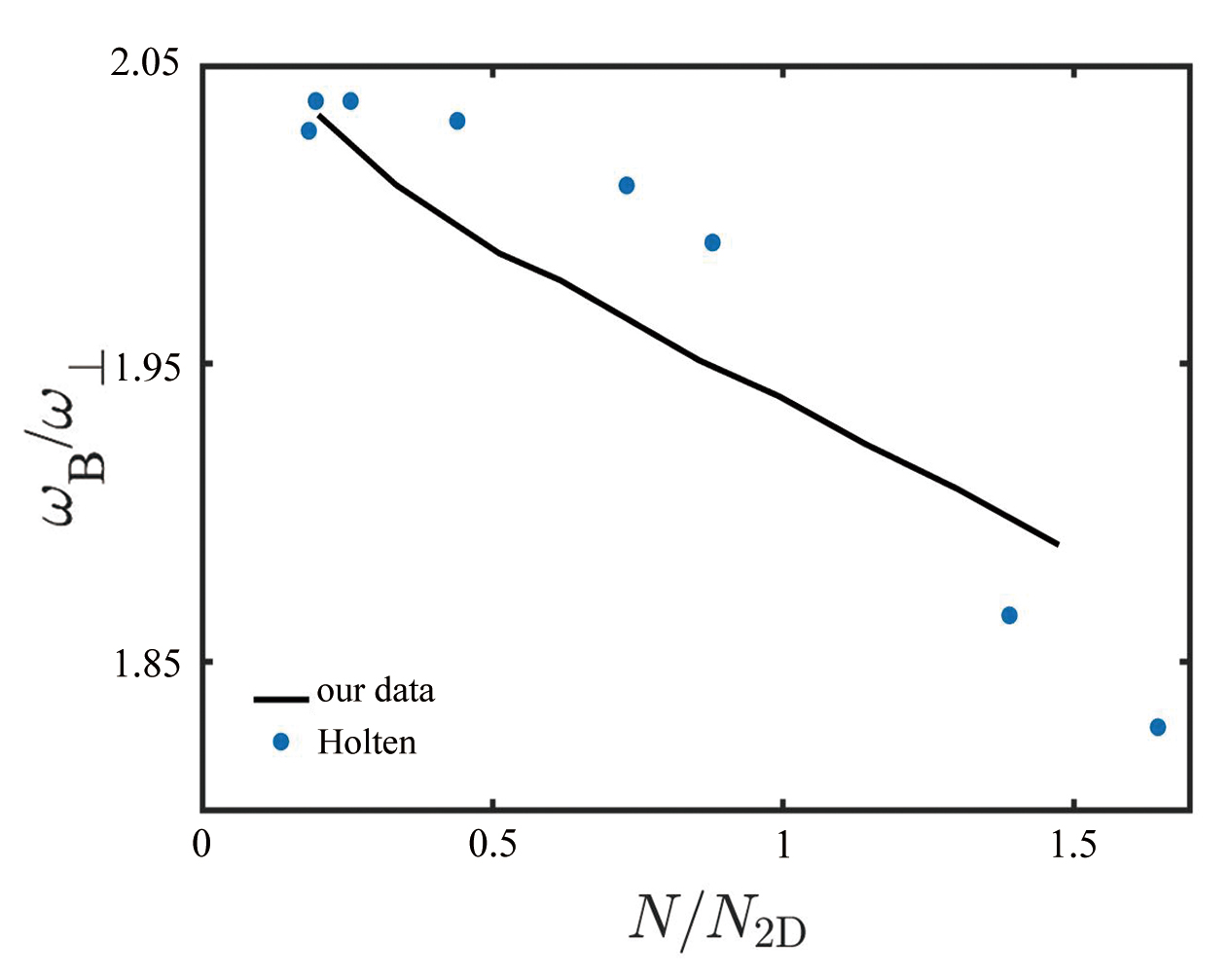}
\caption{(color online) The breathing mode frequency change with the number of the fermions in the trap. Here, the black solid line is the result of our dressed molecules theory, the blue dot is the experimental data of the Holten's group at $0.10\sim0.18T_F$\cite{Holten2018}. And $N_{\rm2D}\simeq(\omega_z/\omega_\perp)^2$ is the 2D threshold for ideal gas.}\label{fig:BM_N}
\end{figure}

Here $\tau$ is the imaginary time after the replacement $t\rightarrow-i\tau$, $\beta=1/(k_{B}T)$ is the inverse temperature, and the effective Hamiltonionian $H_{\rm eff}$ is the real space form of Eq.(\ref{eff-H}). Through the Gaussian integral, fermions can be directly integrated out and we obtain

\begin{align}
Z=&\int\mathfrak{D}[\Delta^{\dag},\Delta]e^{-{\cal S}_{{\rm eff}}[\Delta^{\dag},\Delta]}\\
{\cal S}_{{\rm eff}}[\Delta^{\dag},\Delta]=&-\int dx\ \Delta^{\dagger}\frac{1}{\cal V_{\rm eff}}\Delta^{\phantom{\dag}}-\ {\rm Tr}\ln G^{-1}[\Delta,\Delta^{\dagger}],
\end{align}
where $x=(\tau,{\bf r})$ and $\int dx=\int_{0}^{\beta}d\tau\int d{\bf r}$, $\Delta$ is the auxiliary field, by using the fovrier transforming, the effective action can be written in the momentum space with dimensionless form ( length unit $a_t$ and energy unit $\hbar\omega_z$). Then we can get ${\cal V_{\rm eff}}\rightarrow V_{\rm eff}(iq_{l},{\bf q})=V_{p}-\alpha_p^{2}/(-iq_{l}+{\bf q}^{2}/4+\delta_{p}-2\mu)$, and the fermion matrix
\begin{align}
G^{-1}[\Delta,\Delta^{\dagger}]=
\begin{pmatrix}
i\omega_{n}-\xi_{k} & \Delta\\
\Delta^{\dag} & i\omega_{n}+\xi_{k}
\end{pmatrix}\nonumber\\,
\end{align}
In mean-field level at zero temperature, $\Delta(x)=\Delta_0=\alpha_b\left<d_0\right>+V_b\sum_{\bf k}\left<a_{-{\bf k},\downarrow}a_{{\bf k},\uparrow}\right>$, the grand potential $\Omega=-(T/V)\ln\mathcal{Z}$ can be evaluated as
\begin{eqnarray}
	\Omega_{\rm MF}=-\frac{\Delta_0^{2}}{V_{\rm eff}(0,0)}+\sum_{\bf k}(\xi_{\bf k}-E_{\bf k}+\frac{\Delta_0^{2}}{k^{2}+1}).
\end{eqnarray}
Here, $\xi_{\bf k}=k^2/2-\mu$, $E_{\bf k}=\sqrt{\xi_{\bf k}^{2}+\Delta^{2}}$. However, the mean field theory is not enough to describe the quantum anomaly, and the quantum fluctuation in the system can not be ignored. To have a more quantitative description, we consider quantum fluctuations around the saddle point by writing $\Delta(x)=\Delta_0+\eta(x)$. An exact analytical treatment of the fluctuation contribution is impossible. In our theory, we use the GPF approximation to calculate the quantum fluctuation, whose completeness has been confirmed by previous work\cite{LH1,Hjs1}. The fluctuation grand potential can be written as:

\begin{eqnarray}
	\Omega_{\rm GF}=\sum_ {q,iq_{l}}ln M_{11} + \frac{1}{2}ln \left[1-\frac{M_{12}^{2}}{M_{11}M_{22}}\right]
\end{eqnarray}
\begin{widetext}
While using renomalization condition, we express $M_{11}$ and $M_{12}$ as :
\begin{align}
	 M_{11}=&-\frac{1}{V_{\rm eff}(iq_{l},{\bf q})}+\frac{1}{2}\sum_{\bf k}\Big\{\frac{2}{k^{2}+1}-\left[\left(1+\frac{\xi_{\bf k}}{E_{\bf k}}\frac{\xi_{\bf k+\bf q}}{E_{\bf k+\bf q}}\right)\frac{E_{\bf k}+E_{\bf k+\bf q}}{(E_{\bf k}+E_{\bf k+\bf q})^{2}+q_{l}^{2}}-\left(\frac{\xi_{\bf k}}{E_{\bf k}}+\frac{\xi_{\bf k+\bf q}}{E_{\bf k+\bf q}}\right)\frac{iq_{l}}{(E_{\bf k}+E_{\bf k+\bf q})^{2}+q_{l}^{2}}\right]\Big\} \\
	M_{12}=& \frac{1}{2 }\sum_{\bf k}\frac{\Delta^{2}}{E_{\bf k}E_{\bf k+\bf q}}\frac{E_{\bf k}+E_{\bf k+\bf q}}{(E_{\bf k}+E_{\bf k+\bf q})^{2}+q_{l}^{2}}
\end{align}
\end{widetext}
In this work, we use the local density approximation(LDA) to calculate the breathing mode frequency. And when the total particle number is fixed, the relation between the total particle number and particle density can naturally be written as.
\begin{eqnarray}
	\mu &=&\mu_{c}-\frac{1}{2}m_f\omega_{\perp}^{2}r^{2},\\
	N&=&\int_{0}^{\infty} d{\bf r}n(\mu).
\end{eqnarray}
Here, $\omega_{\perp}$ is the trap frequency of the harmonic trap in the transverse direction, $m_f$ is the mass of fermion. By using the sumrule approach\cite{BM1,BM2}, we can calculated breathing-mode in the zero-temperture.
\begin{eqnarray}
	\hbar^{2}\omega_{B}^{2}=-2\langle r^{2}\rangle\left[\frac{d\langle r^{2}\rangle}{d(\omega_{\perp}^{2})}\right]^{-1}
\end{eqnarray}
where$\langle r^{2}\rangle=N^{-1}\int_{0}^{\infty} d{\bf r}r^{2}n(\mu)$.
Finally,we find the relation between the ground state energy in the center of the potential and the particle number
\begin{eqnarray}
	\Omega_{\rm MF}+\Omega_{\rm GF}&=&-\frac{Nm_f\omega_{\perp}^{2}}{2\pi}	
\end{eqnarray}

\section{Many-Body Calculation Results}\label{res}

Now we focus on the many-body calculation results. First, let's focus on the results of the region near the  resonance point. In Fig\ref{fig:BM_N}, we plot the breathing mode frequency change with the number of the fermions in the trap compare to the experimental data at $ln(k_Fa_{s})\sim-0.1$\cite{Holten2018}. The equation of states calculation of the quantum Monte Carlo(QMC) at this point  is quite different from the experimental data even for the $N/N_{{\rm 2D}}=0.2$\cite{AFQMC}. Because of that, when  $ln(k_Fa_{s})\sim-0.1$, the corresponding effective strength of interaction $V_{\rm eff}/\hbar\omega_z\sim-15$ (see Fig.\ref{fig:para}), which has exceeded the energy gap between the axial excited states and the axial ground state. This makes the axial excited states' fermions can't be neglect even in the case of the particle number $N/N_{\rm2D}$ equal to $0.2$. With the increase of the particle numbers, the proportion of the axial excited states will increase, the axial excited states will play an important role of the systems, which could make the breathing mode frequency $\omega_B/\omega_\perp$ decrease. As these dressed states represent the three dimensional properties in two dimensional systems, the difference between the experimental data and the calculation of the QMC would become bigger for a larger number of particles case. For our theory, as it covers the three types of states of the system, fermions in the axial ground state, fermions in axial excited states, and Feshbach molecular states, our calculation is in good agreement with the experimental results in the strong interaction region even for a larger number case.

\begin{figure}[tbp]
\includegraphics[width=1\linewidth]{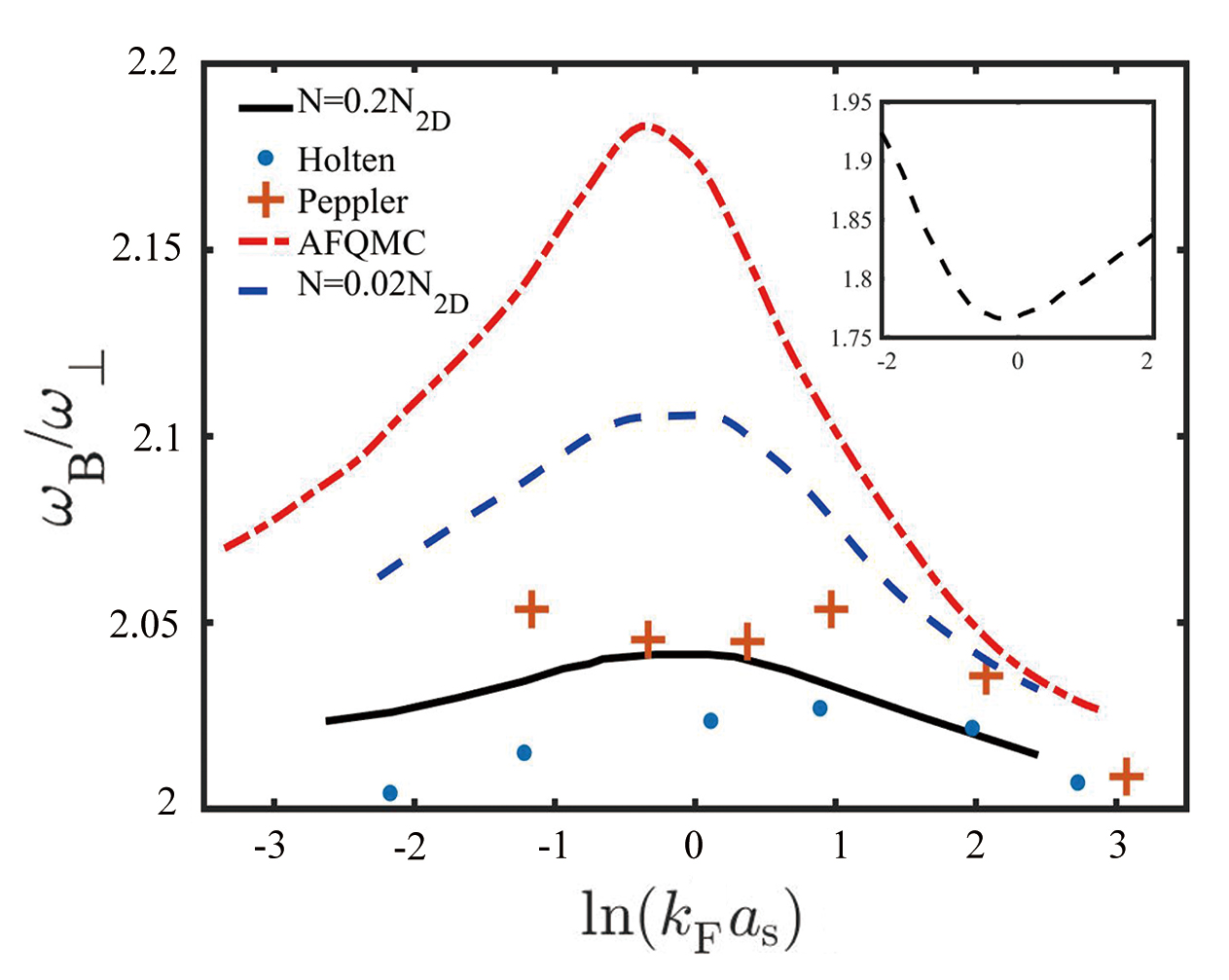}
\caption{(color online) The breathing mode frequency change with the interaction $\ln\left(k_{F}a_{s}\right)$ of the systems at different total number of atoms. The black solid line is the result of our dressed molecules theory, the blue dot is the experimental data of the Holten's group\cite{Holten2018} at $0.10\sim0.18T_F$, the red cross is the experimental data of the Peppler's group at $0.14\sim0.22T_F$\cite{Peppler2018}, for these three date, the total number $N/N_{\rm 2D}\sim0.2$. And the red dot-dashed line is the auxiliary-field quantum Monte Carlo (AFQMC) simulations' data for $N/N_{\rm 2D}\rightarrow0$\cite{AFQMC}, the blue dash line is the result of our dressed molecules theory for $N/N_{\rm 2D}=0.02$. The inset picture is the mean-field calculation of our dressed molecules model at $N/N_{\rm 2D}=0.2$.}\label{fig:BM}
\end{figure}

 In Fig.\ref{fig:BM}, the breathing mode frequency changing with the interaction of the systems compare to the experimental data and the QMC calculation, we show that our theory is a complete low energy effective theory, can explain the experimental observations of the quantum anomaly well for the whole BCS-BEC crossover, not only in the strong interaction region\cite{Holten2018,Peppler2018,AFQMC}. At the BCS-limit, the system can be considered as a non-interaction two dimensional fermi gas, the breathing mode frequency $\omega_B$ must be twice of $\omega_{\perp}$, which is the classcal conclusion for two dimensional system, this phonomenon can be saw in our results or other data. As the system change from the BCS-limit to the resonance point,  due to the effect of the axial excited states fermions which is explaining above, the divergence between the experimental data and the QMC calculation become larger, can be corrected in our theory.

And when the system enters the BEC part, the critical role of the fermions in the properties of the system is replaced by the Feshbach molecules. This requires that the theory describing the quasi-2D system must include the properties of the Feshbach molecular states. On this condition, the system can be approximately understood as a 3D one component interacting Bose gas in a tight-banding harmonic trapping potential. This 3D molecular scattering length is $a_{3D}^m\simeq0.6a_{3D}$\cite{LH1}. The energy shift between those bosen gases and the original Fermi system is $E_{\rm s}\sim E_b$. In BEC region, this energy shift can not be neglected. It makes the effective models, which base on the vacuum energy of the fermi system, not applicable in this case. For our dressed molecular theory, both the physics of 3D Feshbach molecular states and the effect of the energy shift are fully considered in the dressed molecular model, so it can still describe the quasi two-dimensional system completely in the BEC region. In the BEC-limit, it is equivalent to a quasi-2D weakly interacting BEC, the effective 2D molecular scattering length is
\begin{align}
	a_{2D}^m=\sqrt{\frac{\hbar^2\pi}{2m_f\omega_z\mathcal{B}}}\exp\left[-\sqrt{\frac{\hbar^2\pi}{4m_f\omega_z}}\frac{1}{a_{3D}^m}\right]
\end{align}
with $\mathcal{B}\simeq0.905$\cite{Petrov1}. We can also get it from the Eq.(\ref{eq:T3D}). And for our dressed molecular model, as discussed before, it can describe the physics around the bound state energy completely. At this limit, $a_{3D}\rightarrow0^{+}$, $a_{2D}^m\rightarrow0^{+}$, the interaction energy is far less than the trap energy level, so the property of the system is a weak-interaction 2D bose gas, the breathing mode frequency $\omega_B/\omega_{\perp}$ will tend to 2. this phenomenon can be found from Fig.\ref{fig:BM}, especially our mean field result. Therefore, we give a theory which can completely describe the quantum anomaly of breathing mode in the whole BCS-BEC crossover process in the experimental system. We also plot $N/N_{2D}=0.02$ in Fig\ref{fig:BM}. Even in this case, there is still a big difference between our prediction and the dimensional QMC data. This shows that the experimental system can not be described by pure two-dimensional theory.

\section{SUMMARY}\label{Sum}
We have develop a complete theory to describe the two-dimensional quantum anomaly of breathing mode in the experimental system. Within our effective theory, the dressed molecules have been used to characterize the axial excited states' fermions and the Feshbach molecules. It makes our theory describe the high dimensional effect in the strong interaction region and the Feshbach molecular properties in the BEC part well. On the other hand, with the aid of a beyond mean-field, Gaussian pair fluctuation theory, the quantum anomalous properties of the system can be well explained. These make our theory a complete theory describing these quantum anomalies in the two-dimensional experimental systems. Our establishment of the dressed molecules theory for 2D fermions is crucial to understand the conformal anomaly in the quasi low-dimensional quantum systems, and pave the way to investigate the quantum physics in other quasi low-dimensional many-body systems.

\begin{acknowledgments}
We thank Hui Hu, Wei Zhang, Ke-Ji Chen for valuable discussions.
F. W is supported by National Youth Science Foundation of China (CN) (Grand No. 12204105), Educational Research Project for Young and Middle-Aged Teachers of Fujian Province (CN) (Grand No. JAT210041) and Natural Science Foundation of Fujian Province (CN) (Grand No. 2022J05116). L. He is supported by National Key R$\&$D Program (CN) (Grants No. 2018YFA0306503).
\end{acknowledgments}

\begin{widetext}
\appendix

\section{Quasi-2D bubble function}
Here, we will prove that the two particle bubble function for the three dimensional two-component Fermi gases with s-wave interaction by a one-dimensional z directional harmonic trap in this paper is equivalent to that in the Petrov's work\cite{Petrov1}. From Eq.(\ref{eqn:etaseqn}),

\begin{align}
S_p (E)&=\sum_{m,n,{\bf k}} \gamma_{mn}^2[{\cal E}_{m,n, {\bf k}}^{-1}]+\sum_{{\bf k}}\frac{1}{2\epsilon_{\bf k}}\nonumber\\
&=\sum_{n,i=(1\sim n),{\bf k}}\frac{[2-\delta(i)](2n-1)!!^2}{2^{2n}(2\pi)^{1/2}(n+i)!(n-i)!}\frac{1}{E-2\epsilon_{\bf k}-(1+2n)}+\sum_{{\bf k}}\frac{1}{2\epsilon_{\bf k}}.
\end{align}
By using the formula:
\begin{align}
\sum_{i=0}^N\frac{[2-\delta(i)](2N)!}{(2)^{2N}(N+i)!(N-i)!}=1,
\end{align}
the Quasi-2D scattering amplitude can be write as
\begin{align}
S_p (E)&=\sum_{n,{\bf k}}\frac{(2n-1)!!^2}{(2n)!(2\pi)^{1/2}}\frac{1}{E-2\epsilon_{\bf k}-(1+2n)}+\sum_{{\bf k}}\frac{1}{2\epsilon_{\bf k}}\nonumber\\
&=\lim_{N\rightarrow\infty}\frac{1}{2(2\pi)^{3/2}}\{\sum_{n=1}^N\frac{(2n-1)!!}{(2n)!!}ln(n+1/2-E)-2\sqrt{\frac{N}{\pi}}(ln(N)-2)\}.
\end{align}
This is the Quasi-2D bubble function in the Petrov's work\cite{Petrov1}. Here, $S$ is the quantum area, $m$ is the mass of the fermion. To complete this proof, we need to apply the following formula
\begin{align}
\sum_{n=0}^{N}\frac{(2n-1)!!}{(2n)!!}=\frac{2\Gamma(N+3/2)}{N!\sqrt{\pi}}.
\end{align}
The proof is as follows. For $N=0$ case, we can get:
\begin{align}
\sum_{n=0}^{0}\frac{(2n-1)!!}{(2n)!!}=1=2\frac{1}{2}\frac{\sqrt{\pi}}{\sqrt{\pi}}=2\frac{1}{2}\frac{\Gamma(1/2)}{\sqrt{\pi}}=2\frac{\Gamma(N+3/2)}{N!\sqrt{\pi}}.
\end{align}
So if for the $N=M-1$ case, we have $\sum_{n=0}^{N}\frac{(2n-1)!!}{(2n)!!}=\frac{2\Gamma(N+3/2)}{N!\sqrt{\pi}}$, and for the $N=M$ case,
\begin{align}
\sum_{n=0}^{M}\frac{(2n-1)!!}{(2n)!!}&=\frac{2\Gamma(M+1/2)}{(M-1)!\sqrt{\pi}}+\frac{(2M-1)!!}{(2M)!!}\nonumber\\
&=\frac{2\Gamma(M+1/2)}{(M-1)!\sqrt{\pi}}+\frac{\Gamma(M+1/2)}{\sqrt{\pi}(M)!}=\frac{2\Gamma(M+1/2)}{(M)!\sqrt{\pi}}(M+\frac{1}{2})\nonumber\\
&=\frac{2\Gamma(M+3/2)}{(M)!\sqrt{\pi}}.
\end{align}
So we can get the following conclusion, for any $N$, we have $\sum_{n=0}^{N}\frac{(2n-1)!!}{(2n)!!}=\frac{2\Gamma(N+3/2)}{N!\sqrt{\pi}}$. And we also know that, for $N\gg1$ case, $\frac{\Gamma(N+3/2)}{N!\sqrt{N}}\approx1$. Combining these two equation, we can get the following conclusion
\begin{align}
\lim_{N\rightarrow\infty}\sum_{n=0}^{N}\frac{(2n-1)!!}{(2n)!!}=\frac{2\Gamma(N+3/2)}{N!\sqrt{\pi}}=\lim_{N\rightarrow\infty}2\sqrt{\frac{N}{\pi}}.
\end{align}

\end{widetext}


\begin{thebibliography}{99}

\bibitem{QA} William A. Bardeen, Phys. Rev. {\bf 184}, 1848 (1969).

\bibitem{Hofmann} Johannes Hofmann, Phys.Rev.Lett. {\bf 108}, 185303 (2012).

\bibitem{Randeria1989}M. Randeria, J.-M. Duan, and L.-Y. Shieh, Phys.
Rev. Lett. {\bf 62}, 981 (1989).

\bibitem{Berezinskii1972}V. L. Berezinskii, Sov. Phys. JETP {\bf 34},
610 (1972) {[}Zh. Eksp. Teor. Fiz. {\bf 61}, 1144 (1971){]}.

\bibitem{Kosterlitz1973}J. M. Kosterlitz and D. J. Thouless, J.
Phys. C {\bf 6}, 1181 (1973).

\bibitem{Levinsen2015}J. Levinsen and M. M. Parish, Strongly Interacting
Two- Dimensional Fermi Gases, in \textit{Annual Review of Cold Atoms
and Molecules} (World Scientific, Singapore, 2015), Volume 3, Chapter
1, Pages 1-75.

\bibitem{Turlapov2017}A. V. Turlapov and M. Y. Kagan, J. Phys.:
Condens. Matter {\bf 29}, 383004 (2017).

\bibitem{Makhalov2014}V. Makhalov, K. Martiyanov, and A. Turlapov, Phys. Rev.
Lett. \textbf{112}, 045301 (2014).

\bibitem{Martiyanov2016}K. Martiyanov, T. Barmashova, V. Makhalov,
and A. Turlapov, Phys. Rev. A {\bf 93}, 063622 (2016).

\bibitem{Fenech2016}K. Fenech, P. Dyke, T. Peppler, M.G.
Lingham, S. Hoinka, H. Hu, and C.J. Vale, Phys. Rev. Lett. {\bf 116}, 045302
(2016).

\bibitem{Boettcher2016}I. Boettcher, L. Bayha, D. Kedar, P.A.
Murthy, M. Neidig, M.G. Ries, A.N. Wenz,
G. Zrn, S. Jochim, and T. Enss, Phys. Rev. Lett. {\bf 116},
045303 (2016).

\bibitem{Frohlich2011}B. Frohlich, M. Feld, E. Vogt, M. Koschorreck,
W. Zwerger, and M. Kohl, Phys. Rev. Lett. {\bf 106},
105301 (2011).

\bibitem{Sommer2012}A. T. Sommer, L. W. Cheuk, M. J. H. Ku, W. S.
Bakr, and M. W. Zwierlein, Phys. Rev. Lett. {\bf 108}, 045302 (2012).

\bibitem{Zhang2012}Y. Zhang, W. Ong, I. Arakelyan, and J. E. Thomas, Phys. Rev. Lett. {\bf 108},
235302 (2012).

\bibitem{Ries2015}M. G. Ries, A. N. Wenz, G. Zrn, L. Bayha, I. Boettcher,
D. Kedar, P. A. Murthy, N. Neidig, T. Lompe, and S. Jochim,
Phys. Rev. Lett. {\bf 114}, 230401 (2015).

\bibitem{Murthy2015}P. A. Murthy, I. Boettcher, L. Bayha, M. Holzmann,
D. Kedar, M. Neidig, M. G. Ries, A. N. Wenz, G. Zrn, and S. Jochim, Phys. Rev. Lett. {\bf 115},
010401 (2015).

\bibitem{Vogt2012}E. Vogt, M. Feld, B. Frohlich, D. Pertot, M. Koschorreck,
and M. K?hl, Phys. Rev. Lett. {\bf 108}, 070404 (2012).

\bibitem{Holten2018}M. Holten, L. Bayha, A. C. Klein, P. A. Murthy,
P. M. Preiss, and S. Jochim, Phys. Rev. Lett. {\bf 121}, 120401
(2018).

\bibitem{Peppler2018}T. Peppler, P. Dyke, M. Zamorano, S. Hoinka,
and C. J. Vale, Phys. Rev. Lett. {\bf 121}, 120402 (2018).

\bibitem{JH}Johannes Hofmann Phys. Rev. Lett. {\bf 108}, 185303 (2012).

\bibitem{YuZH} Chao Gao and Zhenhua Yu Phys. Rev. A {\bf 86}, 043609 (2012).

\bibitem{WF} F Wu, J Hu, L He, XJ Liu, H Hu, Physical Review A {\bf 101} (4), 043607 (2020).

\bibitem{WZ} R Zhang, F Wu, JR Tang, GC Guo, W Yi, W Zhang, Physical Review A 87 (3), 033629 (2013).

\bibitem{LMD1}J. P. Kestner and L.-M. Duan, Phys. Rev. A {\bf 74}, 053606 (2006).

\bibitem{LMD2}J. P. Kestner and L.-M. Duan, Phys. Rev. A {\bf 76}, 063610 (2007).

\bibitem{WZ1}W. Zhang, G.-D. Lin, and L.-M. Duan, Phys. Rev. A {\bf 77}, 063613 (2008).

\bibitem{WZ2} W. Zhang, G.-D. Lin, and L.-M. Duan, Phys. Rev. A {\bf 78}, 043617 (2008).

\bibitem{WY1} W. Yi and L.-M. Duan, Phys. Rev. A {\bf 73}, 063607 (2006).

\bibitem{Petrov1} D. S. Petrov and G. V. Shlyapnikov, Phys. Rev. A {\bf 64}, 012706 (2001).

\bibitem{Holten}M. Holten, L. Bayha, A. C. Klein, P. A. Murthy, P.M. Preiss, and S. Jochim, Phys. Rev. Lett. {\bf 121}, 120401(2018).

\bibitem{Peppler}T. Peppler, P. Dyke, M. Zamorano, S. Hoinka, and C. J. Vale, Phys. Rev. Lett. {\bf 121}, 120402 (2018).

\bibitem{WF1}F Wu, J Hu, L He, XJ Liu, H Hu, Phys. Rev. A {\bf101} , 043607 (2020).

\bibitem{Hjs1}J Hu, F Wu, L He, XJ Liu, H Hu, Phys. Rev. A {\bf101} , 013615 (2020).

\bibitem{BM1}C.Menotti, S.Stringari, Phys. Rev. A {\bf66} , 043610 (2002).

\bibitem{BM2}H.Hu, G.Xianlong, XJ Liu, Phys. Rev. A {\bf90} , 013622 (2014).

\bibitem{QC}Q. Chen, J. Stajic, S. Tan, and K. Levin, Phys. Rep. {\bf 412}, 1 (2005).

\bibitem{LH1}Lianyi He, Haifeng Lv, Gaoqing Cao, Hui Hu, Xia-Ji Liu, Phys. Rev. A {\bf 92}, 023620 (2015.)

\bibitem{AFQMC} H. Shi, S. Chiesa, and S. Zhang,
Phys. Rev. A {\bf 92}, 033603 (2015).

\end{thebibliography}
\end{document}